\documentclass{aa}  
\usepackage{linenoaa}
\usepackage{graphicx}
\usepackage{txfonts}
\begin{document} 
   \title{Compact near-infrared sources in the center of the extraordinary galaxy IC~860}
      \author{J. S. Gallagher
          \inst{1,2}
          \and
          L. Schisgal\inst{1}
          \and 
          G. C. Privon\inst{3,4,5}
          \and
          S. Aalto\inst{6}
          \and
          S. K\"{o}nig\inst{6}
          \and
          R. Kotulla\inst{2}
          \and
          J. Mangum\inst{3}
          \and
          W. St. John\inst{1}.
          \and
          D. Rigopoulou\inst{7,8}
          \and
          K. Alatalo\inst{9,10}
       }   

   \institute{Department of Physics and Astronomy, Macalester College,
                1600 Grand Ave, St. Paul, MN 55438 USA \\
              \email{jgallag1@macalester.edu}
         \and
              Department of Astronomy, University of Wisconsin-Madison, 475 N. Charter St., Madison, WI 53706 USA
         \and
             National Radio Astronomy Observatory, Edgemont Rd.
             Charlottesville, VA 22903 USA
        \and 
             Department of Astronomy, University of Florida, P.O. Box 112055, Gainesville, FL 32611 USA
        \and
            Department of Astronomy, University of Virginia, P.O. Box 400325, Charlottesville, VA 22904 USA
        \and
            Department of Space, Earth and Environment, Onsala Space Observatory, Chalmers University of Technology, 43992 Onsala, Sweden  
         \and
         Department of Physics, University of Oxford, Keble Road, Oxford OX1 3RH, UK 
         \and
         School of Sciences, European University Cyprus, Diogenes street, Engomi, 1516 Nicosia, Cyprus
         \and
 	Space Telescope Science Institute, 3700 San Martin Drive, Baltimore, MD 21218, USA
	\and
	William H. Miller III Department of Physics and Astronomy, Johns Hopkins University, Baltimore, MD 21218, USA             }

   \date{Received} 
  \abstract
   {Hubble Space Telescope (HST) images are used to study the structure of the central regions of the luminous infrared galaxy (LIRG) IC~860. IC~860 is of special interest as a system with an extreme central concentration of molecular gas cloaking its compact obscured nucleus (CON). The CON provides most of the 1.5 $\times$ 10$^{11}$~L$_{\odot}$ luminosity in IC~860 from an undetermined combination of stellar and active galactic nucleus power sources.}
   {We mapped and photometered the central molecular zone (CMZ) of IC~860 motivated by a previous detection of a luminous compact nuclear source. Our objective was to study the properties of the CMZ and its relationship to the CON, which we identified as an opaque central region in archival near-infrared (NIR) images.}
   {We measured the coordinates of the compact NIR source, IC860-a, from HST coordinates calibrated with Gaia positions.  The photometry of the HST images yielded magnitudes, colors, and high V-band dust optical depths based on foreground screen dust models. Photometry corrected for dimming by dust yielded NIR luminosities of IC860-a and the CMZ.}
   {IC~860 has distinct compact central luminosity sources. Most of the NIR luminosity is from the CMZ, while the LIRG-CON is an obscured region centered in the CMZ. IC860-a, on the northeastern side of the CMZ, is offset by $\approx$0.2\arcsec from the CON, has a luminosity of $\sim 10^9$~L$_{\odot}$, and may be a massive young stellar complex or intruding nucleus.} 
   {The inner CMZ in IC~860 contains two luminous compact objects. The CON is identified for the first time as an obscured central source, while the structure of the CMZ+CON is complicated by the presence of IC860-a, compact object and possibly a massive young stellar system or second nucleus.  The presence of IC~860-a in combination with the CON is a further signature of the unusual evolution of the gas-rich IC~860 CMZ.}


   \maketitle
   \nolinenumbers

\section{Introduction}

IC~860 is an SBa galaxy located at a distance of 59~Mpc. Its unusual nature was noticed when \cite{Schmelz86} detected 18~cm OH absorption. Soon thereafter, IC~860 was recognized to be a luminous infrared galaxy (LIRG) \citep{Soifer87,Carico88},  \cite{Mirabel88} found an unusual global H\rm{I} 21~cm line in absorption, and \cite{Kazes88} set a path for future studies by describing IC~860 as a very peculiar galaxy. \cite{Condon90} found  a compact central radio source  $<$0.5\arcsec in size. The intrinsic nature of this galaxy and its sources of infrared luminosity, however, remained unclear, but most investigators considered it to be either  a non-interacting galaxy or a possible product of a minor merger \citep[see discussions in][]{Baan17,Luo22}. IC~860 has also  been classified as a post-burst galaxy with large-scale optical emission from shocks that suggest the presence of an active galactic nucleus (AGN) that is yet to be confirmed \citep[][and Figure~\ref{fig:ic860_opt}]{Alatalo16,Luo22}.

Initial interferometric millimeter observations with modest angular resolution did not reveal the presence of a powerful nucleus in IC~860 \citep[e.g.,][]{Imanishi06}, but showed an inner molecular disk \citep[see, e.g.,][and references therein]{Alatalo24}. Mid-infrared spectroscopy of the center of IC~860 with the {\it Spitzer Space Telescope} and radio observations of ammonia lines, however, detected absorption from molecules located in dense hot molecular matter, demonstrating that unusual conditions exist \citep{Lahuis07,Mangum13}. Signatures in the form of maser emission and outflows further suggested that an obscured source of central activity existed, as discussed by  \cite{Baan06}, among others, who also provided evidence for the presence of heavy dust obscuration toward compact sources in the center of IC~860 with their subarcsecond radio continuum imaging. Subsequent high angular resolution observations with millimeter/submillimeter interferometers established that the nucleus of IC~860 is its primary luminosity source. The nucleus is a compact highly obscured nucleus (CON). 

A CON is defined as a compact galaxy nucleus displaying vibrational HCN submillimeter emission lines arising in dusty, dense, warm interstellar matter that envelopes the nucleus \citep[see][]{Aalto13,Falstad21}. Compact obscured nuclei also can be distinguished by their extreme opacities \citep{Donnan23} and mid-infrared spectral characteristics \citep{GarciaBernete25}. The radius of a CON is set by wavelength-dependent dust opacity which is low in the millimeter.  The existence of CONs demonstrates that galaxies in the current epoch can collect substantial amounts of low angular momentum molecular gas close to their nuclei. This low angular momentum gas has the potential to the fuel rapid growth of nuclear regions, such as nuclear star clusters and central massive black holes \citep[e.g.,][]{Cen15,Volonteri15,AnglesAlcazar17,RamosAlmeida17,Hickox18,Gorski24,Nishimura24,Kritos26}. That CONs make use of this fuel can be seen from their presence primarily in systems with L$_{FIR} \gtrsim$10$^{11}$~L$_{\odot}$, where they are the  major power sources in their LIRG hosts \citep{Falstad21,Nishimura24}. Unfortunately, the extreme optical depths in CONs hide the interior power sources from the X-ray to millimeter bands. Possible sources of energy include active nuclei, starbursts, or accretion luminosity from gas infalling to the CON, or likely some combination of these \citep[e.g.,][]{Gorski23}. 

Before the IC~860 CON was discovered, observations by \cite{AlonsoHerrero06} with the Near Infrared Camera and Mulit-Object Spectrometer 2 (NICMOS2; NIC2) on the Hubble Space Telescope (HST) showed a point source nucleus candidate in IC~860 that is visible in the near-infrared (NIR). However, the absolute position of this source, IC860-a in this paper, was not known, and therefore its relationship to the CON that was discovered later was not explored.  Here we revisit the properties of the central regions of IC~860 with an emphasis on properties of the CON and IC860-a. Our study benefits from the access to archival high angular resolution  images obtained with the HST. Improved coordinate reference frames based on stellar positions from Gaia in Wide Field Camera 3 (WFC3) HST images now allow us to match positions from HST images to those obtained with millimeter/submillimeter interferometers at subarcsecond levels.   

The objective of this study is to provide an HST context for JWST observations of the center of IC~860 obtained in  program JWST GO-01991, P.I. G. Privon. In the next section we review the archival HST observations that provide the data for our study and for the first time reveal the presence of an opaque CON in the NIR. Properties of the CMZ and inner 300~pc  are considered in \S\ref{sec:centralstruct}. The coordinates of IC860-a are tabulated in \S\ref{sec:nucpos} and compared with those of the CON, which establishes that IC860-a is not the CON. Section~\ref{sec:northobjptm} presents photometry from WFC3 and NIC2 images. Section~\ref{sec:ssctrdust} considers the levels of central dust obscuration in IC~860 derived from optical--NIR color maps.  The dust corrected photometry leads to lower bound estimates for source luminosities. We discuss the results in \S\ref{sec:discuss} and give our conclusions in \S\ref{sec:conclude}.\

\section{Observations}\label{Sec:obs}

\begin{figure}
   \centering
   \includegraphics[width=0.25\textwidth]{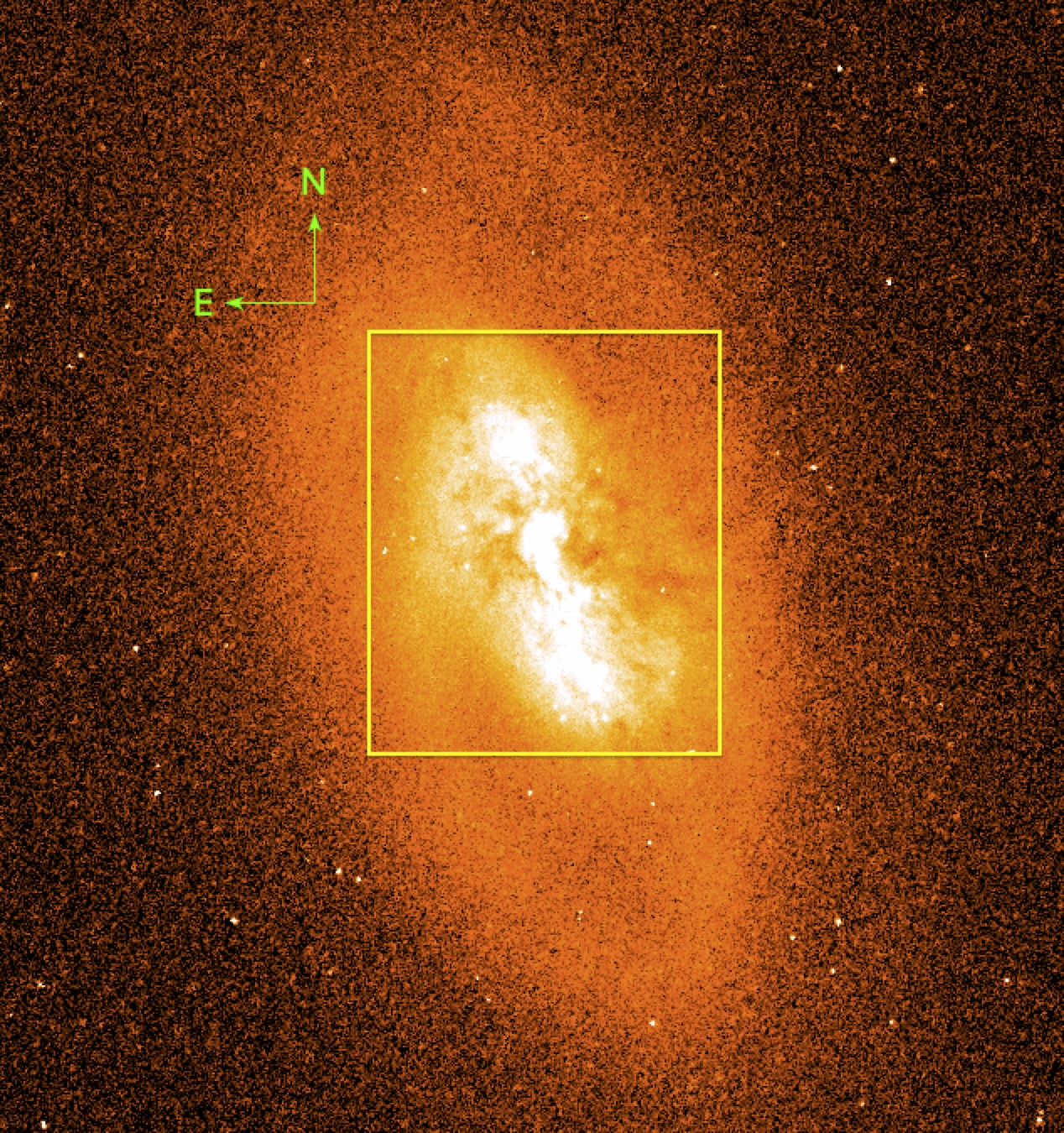} \
   \includegraphics[width=0.25\textwidth]{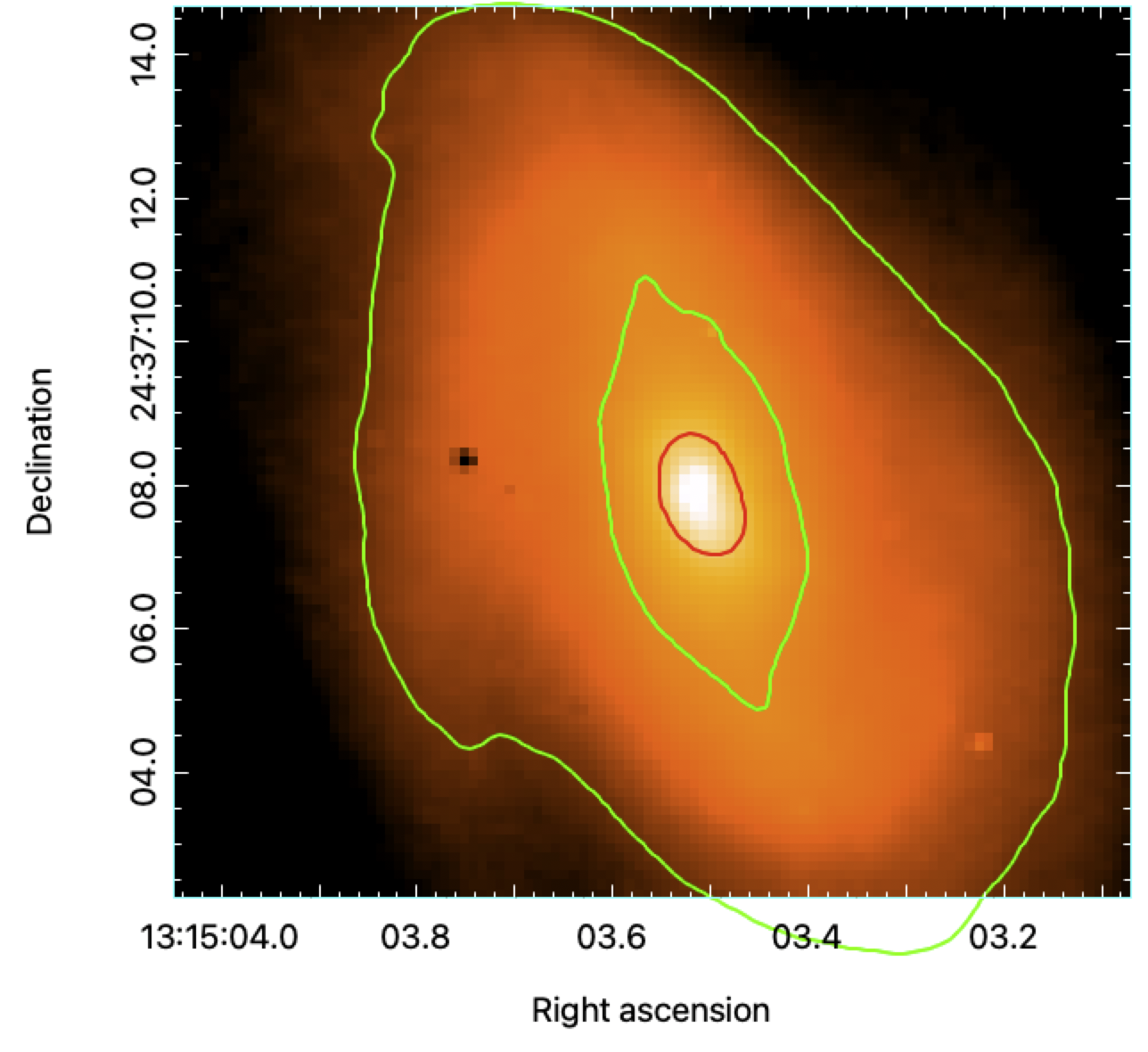}
      \caption{{\it Upper}: HST WFC3/UVIS F438W blue light image of IC~860.  This shows the asymmetric dominant spiral arms and inner star-forming bar with a bright but dusty center. \cite{Luo22} suggest that the strong spiral arm is a possible signature of a past interaction.  The 15$\times$ 15\arcsec$^2$ (4.3 $\times$ 4.3~kpc$^2$) box defines the region shown in the lower image. The central dust lanes extending on the west side of center are typical of extraplanar material emanating from the position of the CON and its immediate surroundings (see Fig.~\ref{fig:dustdepth}). The bright spots along the major axis appear to be regions of low obscuration, possibly containing younger stars in an inner bar structure. {\it Lower:} Central region of IC~860 from the WFC3/IR F140W observation of the boxed region in the upper figure. Isophotes are included to outline the structure of the inner galaxy while the central red contour delineates the high infrared surface brightness CMZ of IC~860 that contains IC860-a and is the location of its centrally concentrated CO 1-0 emission \citep{Alatalo24}.}
         \label{fig:ic860_opt}
   \end{figure}

Our study is based on HST archival optical and near-infrared (NIR) images of IC~860 obtained with the WFC3.\footnote{HST Archival WFC3 images are from programs GO-10169 P. I. Alonso-Herrero (NIC2); datasets n8zl3010-040  and GO-14715, P.I. K. Alatalo (WFC3); dataset id8n0101, id8n02010-030, id801q9Q,  and id8n02wpq retrieved from the Mikulski Archive for Space Telescopes}  An overview of IC~860 from the HST imaging is shown in Figure~\ref{fig:ic860_opt}. This system has well-defined spiral arms and a linear bar, consistent with its SBa morphological type. 

The high intensity central region, shown by the contour in Figure~\ref{fig:ic860_opt}, is prominent in the NIR but does not stand out in optical images. The gas-rich nature of the center of IC~860, however, can be seen in the form of extensive dust features to the west side of the galaxy's center. This pattern indicates that dusty gas is above the galaxy's mid-plane and only appears in absorption when observed against the stellar disk. Dusty extraplanar gas on the west side is below the disk and not detectable in absorption. The region within the inner contour in Figure~\ref{fig:ic860_opt} and to the west of the inner extraplanar dust features is the target of our investigation. 

Photometric and source position measurements were obtained from the reduced fits images with corrections for charge transfer inefficiencies (fits drc file types) provided by the Mikulski Archive for Space Telescopes (MAST). Photometric calibration came from parameters in the image headers using standard techniques to convert the data count rates per pixel to magnitudes and fluxes. We made initial position measurements using image coordinates included with the MAST data that we finalized by re-reducing one image. 

\section{Overview of central structures}\label{sec:centralstruct}

\begin{figure}
  \begin{centering} 
        \includegraphics[width=0.15\textwidth]{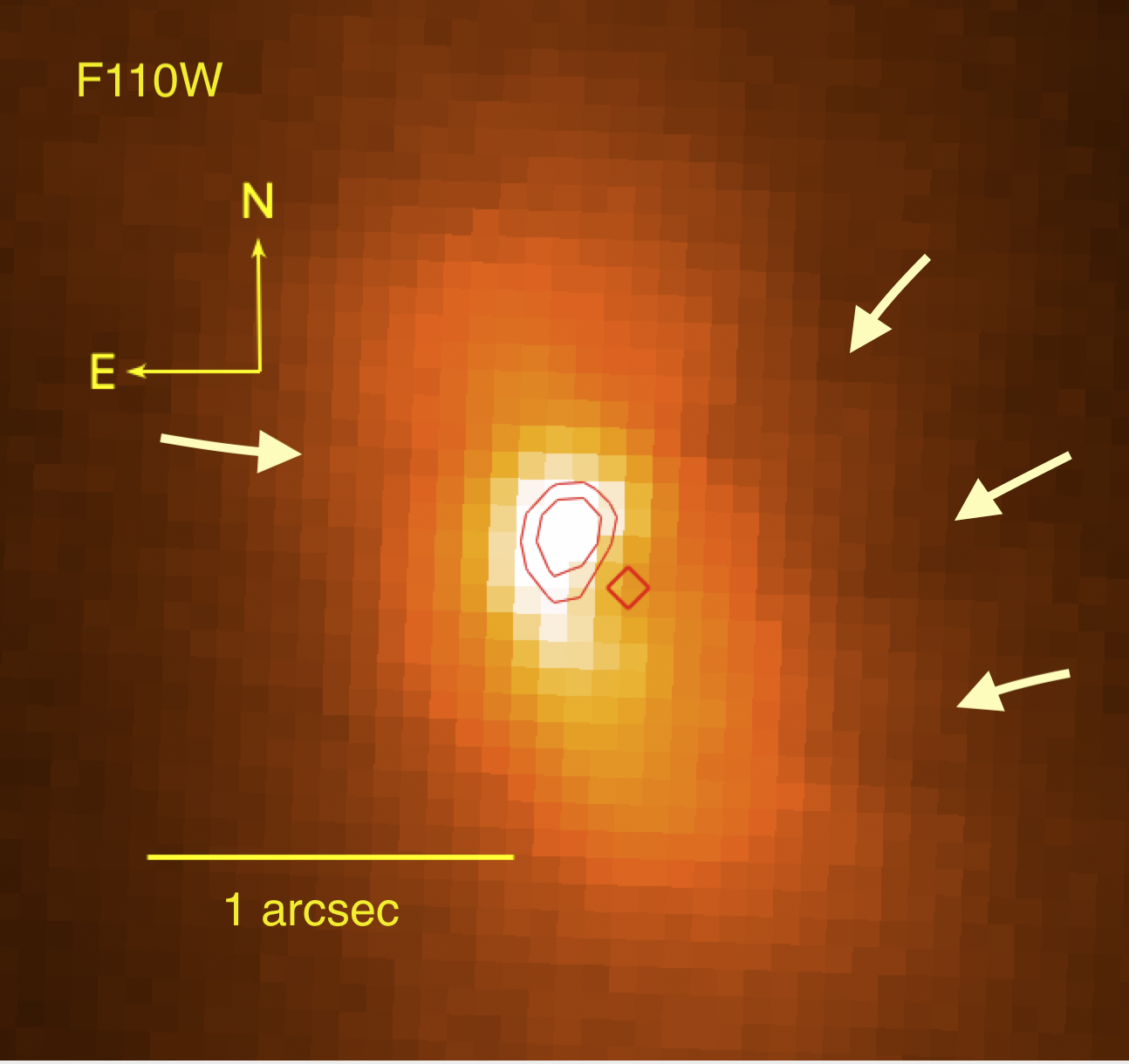}
        \includegraphics[width=0.15\textwidth]{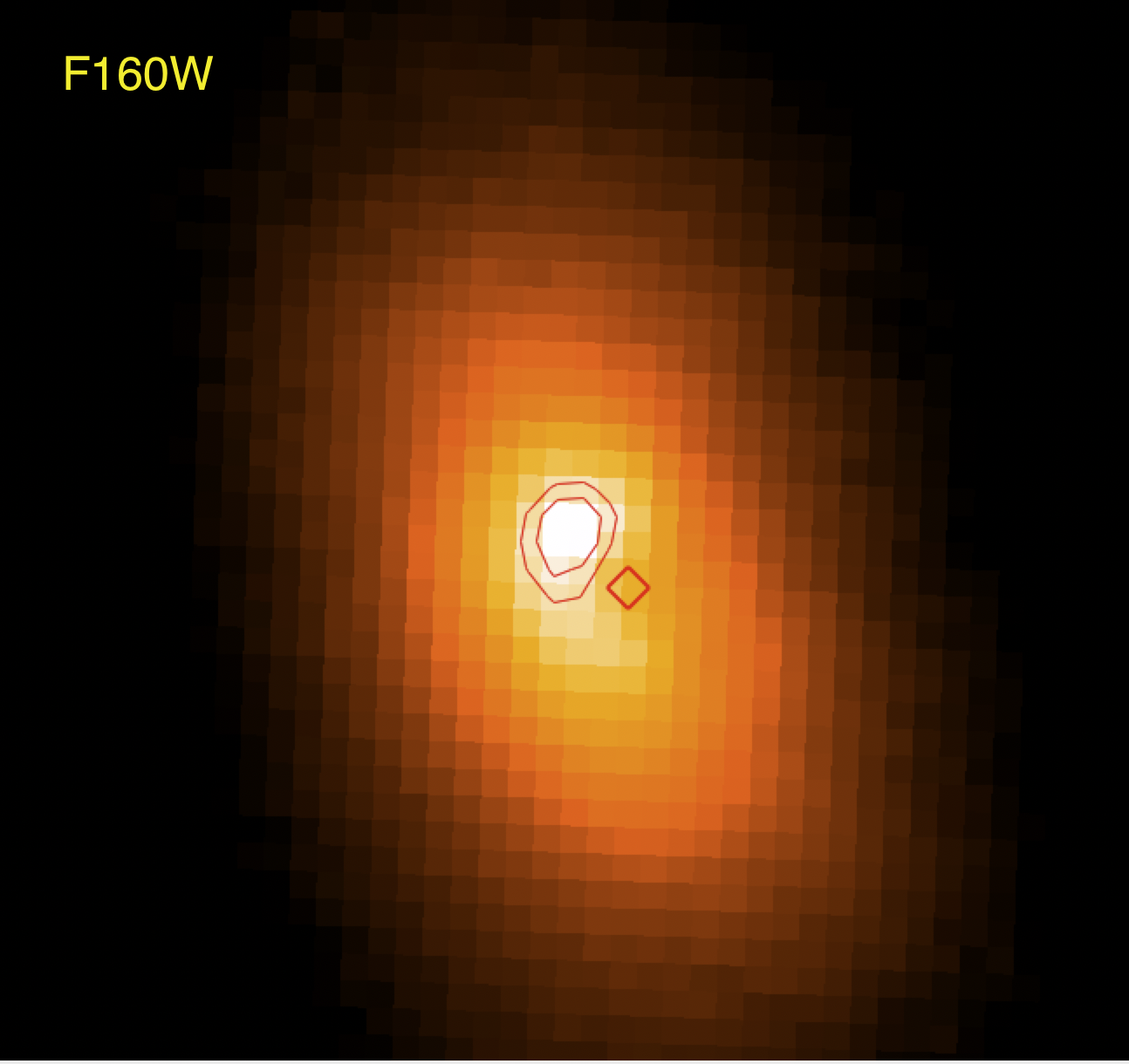} \\
        \includegraphics[width=0.15\textwidth]{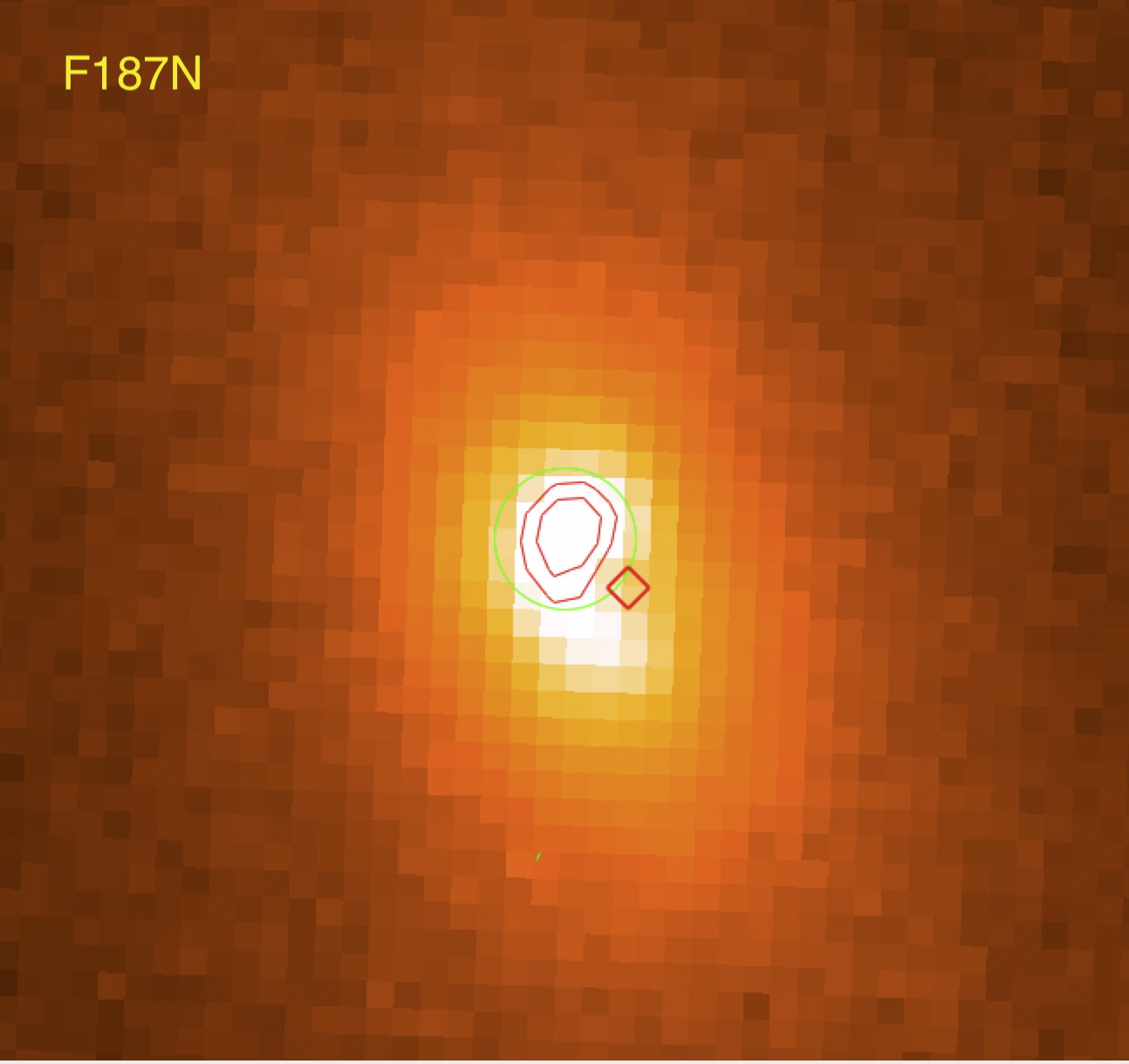}
        \includegraphics[width=0.15\textwidth]{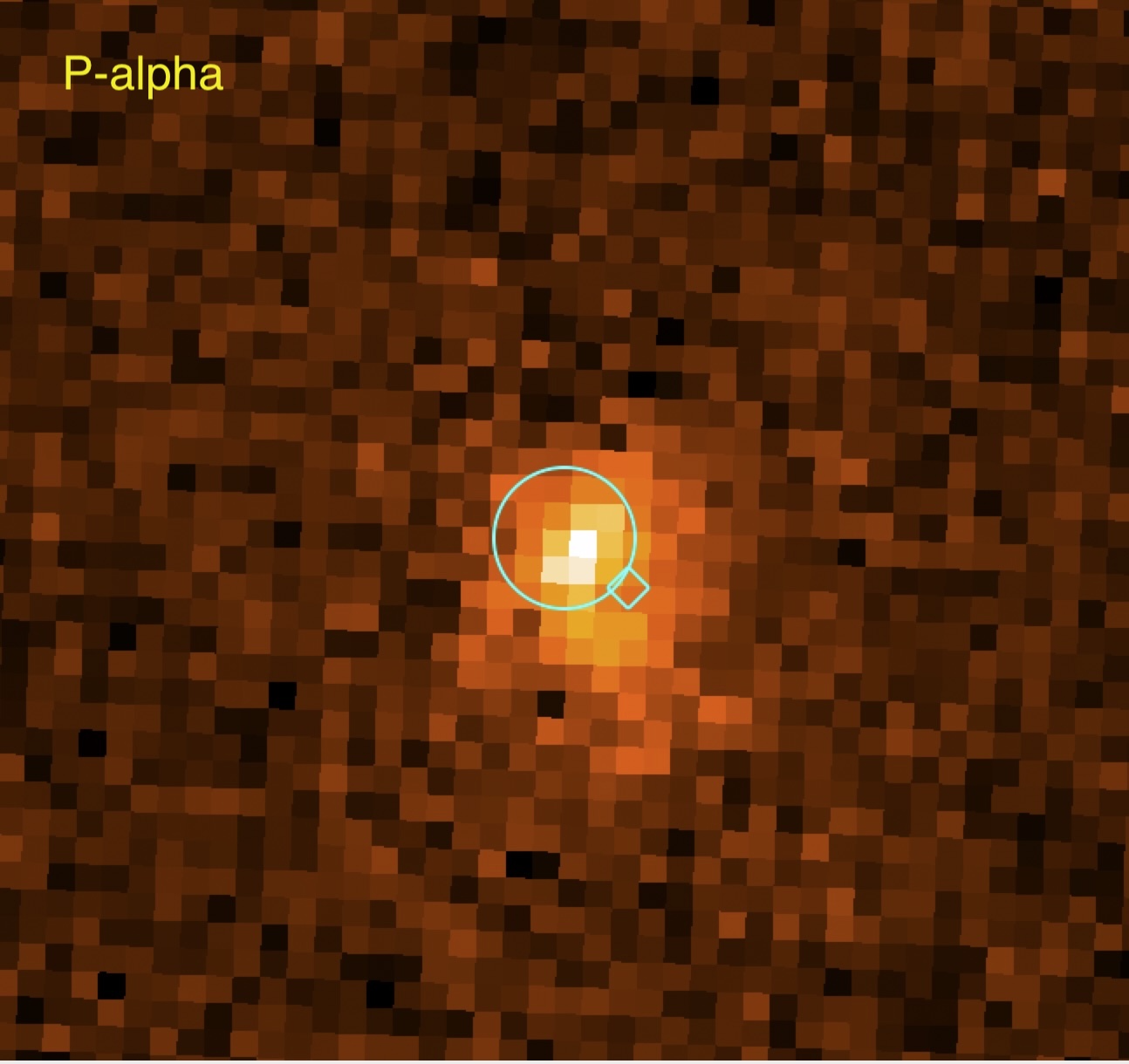}
    \caption{Archival NIC2 images of the center of IC~860. The arc of enhanced surface brightness outlining the CMZ and IC860-a (contours) and the CON absorption region (diamond) are the main NIR features within this zone. Dust absorption clouds (arrows) are faintly visible in the F110W image as darker regions around IC860-a and the position of the CON. The F187N is a continuum band at the IC~860 redshift. The CON appears as a dark source in the continuum images, thus indicating a high NIR optical depth. The Pa$\alpha$ image includes a circle at the position of IC860-a and shows H-recombination emission from much of the CMZ. Coordinates are not included since the NIC2 data reference frames have coordinate offsets.}
    \label{fig:nic2_cmz}
  \end{centering}
\end{figure}

The center of IC~860 is obscured by dust in optical bands (see Figure~\ref{fig:ic860_opt} and \cite{Luo22}) but emerges in NIR images. Figure~\ref{fig:nic2_cmz} shows the high surface brightness center of the oval region marked in Figure~\ref{fig:ic860_opt} (lower) in NIC2 NIR images. The inner part of the bar contains a zone of enhanced surface brightness with a major axis of 1.2\arcsec and a minor axis of $\approx$0.4\arcsec (340 $\times$ 115~pc$^{2}$). The major axis at PA=18$^{\circ}$ aligns with the bar of IC~860.  We considered this region to be the IC~860 CMZ in this paper. Longer wavelengths minimize the influence of dust obscuration on the central morphology.  The central zone of enhanced NIR surface brightness corresponds to molecular emission exterior to the CON \citep{Aalto15,Baan17,Luo22,Alatalo24} and we identify it with the central molecular zone (CMZ) of IC~860. 

The NIC2 images show the inner part of the CMZ as a luminous curved arc with IC860-a appearing as a compact source near its northeastern edge. These data have the advantage of small pixels at the cost of a reduced field of view relative to the WFC3/IR channel.  Submillimeter observation measure R$_{submm}  \sim$10~pc (R$_{obs} (\approx$ 0.03~\arcsec) for the CON in IC860. These data show the CON is the main luminosity source and IC~860  and that the inner, high density core of the CON's interstellar envelope that has a mass of $\lesssim$10$^8$~M$_{\odot}$ \citep{Aalto15,Aalto19,GonzalezAlfonso19,Falstad21}. 

The radius of the dark, obscured region at the position of the CON is $\sim$ 1~pixel =0.075\arcsec (20 pc) in NIC2 NIR images. This marks the position of the opaque envelope that we assign to the outer surface of the spheroidal CON and is the first identifcation of a CON in NIR absorption (see \S\ref{sec:nucpos} below).  Dust lanes faintly detected in the NIC2 F110W image connect to the area enclosed by the CMZ arc, the location of the CON, and around the location of IC860-a.  This pattern is suggests the presence of organized gas flows in the central regions of IC~860 as discussed by \cite{Gorski23}.  The CON is the source of a strong  and likely dusty molecular wind which we did not consider when discussing the CON opacity. This issue needs to be explored with better data.

    \section{Positions of IC860-a and the CON}\label{sec:nucpos}

\begin{figure}
   \begin{centering}
      \includegraphics[width=0.48\textwidth]{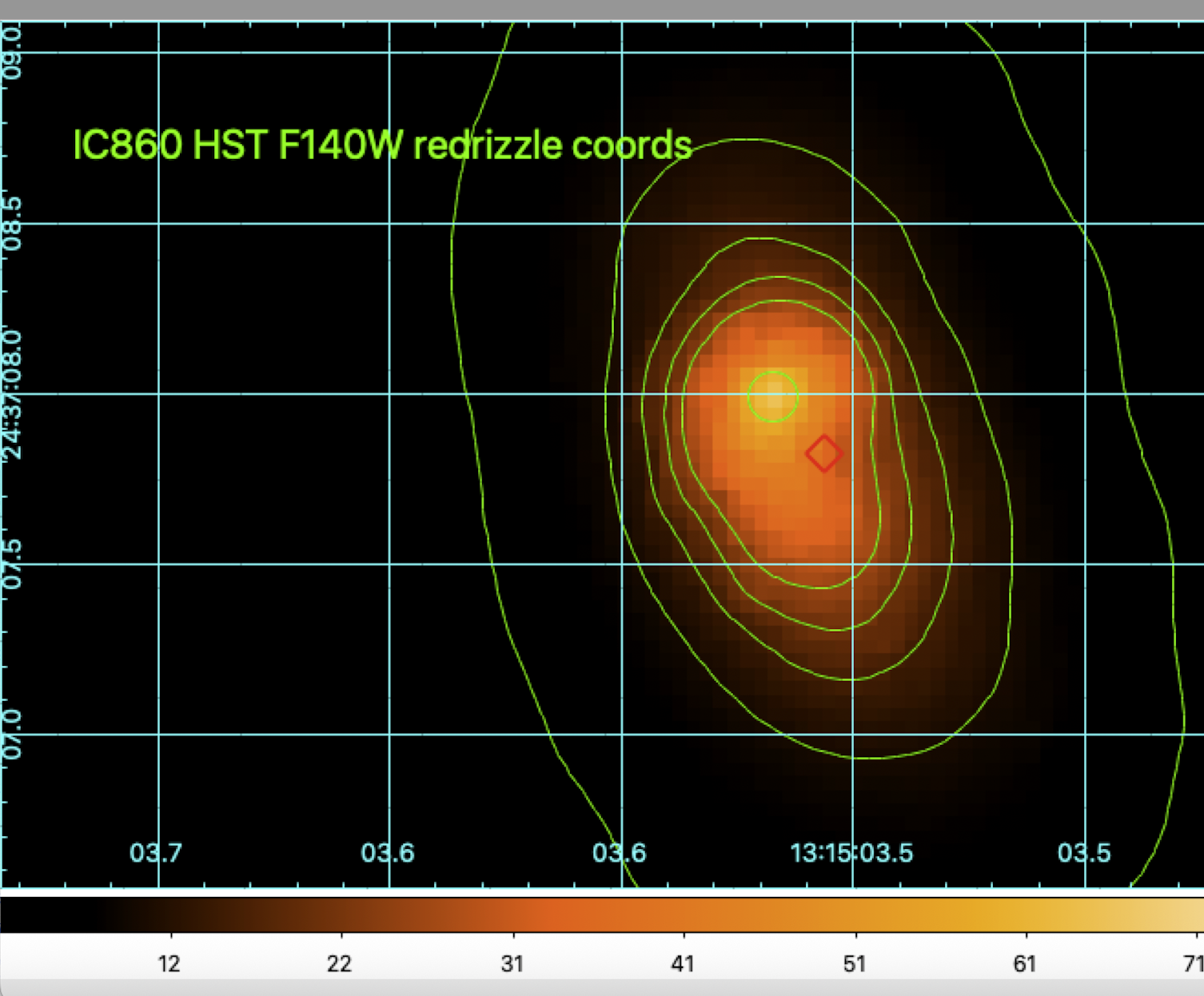} 
      \caption{WFC3 F140W image of the inner IC~860 CMZ region with updated J2000 FK5 coordinates. The locations of   IC860-a (yellow circle) and the CON (red diamond) are displayed in this WFC3 F140W image of IC~860. }
         \label{fig:ic860-a_ctr}
    \end{centering}
\end{figure}

 Our objective in measuring the NIR position of IC860-a was to determine its location relative to that of the CON. The CON's coordinates are accurately known from a combination of radio continuum and molecular line observations \citep[e.g.,][]{Aalto19}. IC860-a is detected in F814W and longer wavelength filters (see Figure~\ref{fig:ic860-a_ctr}).  We searched for  IC860-a by comparing archival WFC3 optical and NIR images with the NIC2 images and established that it is the point source measured by \citep{AlonsoHerrero06}. Our initial reconnaissance also showed IC860-a to be at the same position in WFC3 filters ranging from F814W to F160W.  The absence of IC860-a in the F606W and F475W WFC3 images reflects the high levels of dust obscuration deduced by  Alonso-Herrero {\it et al.}, a point further confirmed by our analysis (see \S~\ref{sec:ssctrdust}).

 We first measured positions for IC860-a in the WFC3 F814W, F140W, and F160W images using the MAST-supplied coordinate frames with the IRAF task radprof. This tool plots intensity and of the region in multiple radial cuts and derives the location of the intensity peak and its positional error in pixel coordinates. The results showed IC860-a to have a well defined intensity profile above the complex background, especially in the NIR filters and that the peak position was consistent in position in the HST filters. Two archival F140W observations of IC~860 taken 6.5 months apart with different orientations and positions on the detector yielded the same coordinates for IC860-a. Due to the small field of view we could not obtain the position of IC860-a from the NIC2 images. 
 
 Although the F814W image offered the advantage of the best angular resolution the effects of dust limit the quality of our measurements of the source location and flux. We chose the WFC3/IR F140W id8n01q9q image  to determine our adopted position of IC860-a. This filter combined reasonable angular resolution and reduced sensitivity to dust. We updated coordinates for the F140W image with positions from the Gaia DR3 catalog matched to stars in the field of view. In so doing we omitted one star in the Gaia catalog that lacked a proper motion measurement since this star had an uncertain position at the time of the F140W observation, and also dealt with issues associated with bright stars. The  corrected image redrizzled to a 0.04\arcsec pixel scale is shown in Figure~\ref{fig:ic860-a_ctr} along with a photocentroid fit to IC860-a with the IRAF task radprof using a range in region sizes.   The resulting scatter at the 0.5 pixel (0.02\arcsec) level in each coordinate led to our total positional uncertainty for IC860-a of $\pm$0.03\arcsec. 
  
Our adopted position for IC860-a from updated coordinates in the F140W image is J2000 13:15:03.515$\pm$0.002, +24:37:07.96$\pm$0.02. This significantly differs from the IC~860 CON's position of J2000 13:15:03.506 +24:37:07.82 that has a total uncertainty of 0.02\arcsec  \citep{Aalto19}. IC860-a is offset 0.16$\pm0.034$\arcsec ($\approx$50~pc in projection) to the north-northeast of the CON in IC~860. The CON in IC~860 is not coincident with IC860-a; these are distinct structures.

As shown in Figure~\ref{fig:ic860-a_ctr}, the CON absorption feature is located at the inner edge of the curved CMZ structure and at the base of the central dusty region extending to the west that may mark a dusty wind. We measured the position of this opaque region in the NIC2 F110W image relative to that of IC860-a and determined that its location agrees with that of the CON. This position places the CON near the center of the CMZ, consistent with its status as the likely nucleus of IC~860. \cite{Skipper18} find that the radio source associated with the CON is possibly slightly displaced from the SDSS measurement of the isophotal center of IC~860, but our HST imaging with the updated coordinate frames confirms that the radio position of the CON is close to the center of the CMZ. A summary of our position measurements for IC860-a is in Table~\ref{tab:positions}.

\begin{table}
    \caption{IC~860 central structure positions.}
    \centering
    \begin{tabular}{lccl}
    \hline\hline
    \\
     Object & RA   & DEC \\  \hline
      {\it  CON}$^a$ & {\it 13:15:03.506} & {\it  24:37:07.812} \\ 
     IC860-a adopted$^b$  & 13:15:03.515$\pm$0.002 &  24:37:07.96$\pm$0.02 \\ 
     IC860-a, F814W$^{c}$ & 13:15:03.514 & 24:37:07.96 \\
     IC860-a Dust$^d$ & 13:15:03.517$\pm$0.002 & 24:37:07.80$\pm$0.02 \\
     \hline
    \end{tabular}
     \label{tab:positions}
     \tablefoot{Positions are J2000 ICRS derived from the WFC3 images, as discussed in the text and below.  {\it Notes:} (a) CON position from \cite{Aalto19} Table~3 with a positional uncertainty of $\pm$0.02\arcsec; (b) Position from re-reduced F140W image in this paper;  (c) From archival image coordinate frame, uncertainty not well determined due to dust obscuration impact on the profile; (d) IC860-a reddening peak position measured from our optical depth maps (see Figure~\ref{fig:dustdepth}).}
\end{table}

\section{Photometry}\label{sec:northobjptm}

\subsection{IC~860a}\label{sec:ssNnucirptm}
 
We first removed the wider galaxy background by subtracting data smoothed by medians of 7$\times$7, 9$\times$9 and 15$\times$15 pixel$^2$ for WFC3/UVIS, WFC3/IR/ and NIC2 images respectively before making photometric measurements.  The IRAF radprof task allowed us to locate the position of the intensity peak of IC860-a, determine a background level, and obtain aperture  photometry 
With this information we adopted an aperture with r$_{ap}$= 0.15\arcsec that corresponds to the approximate level where IC~860-a merges into the CMZ background. Count rates and fluxes for all filters were derived using information in the image headers and standard HST magnitude calibrations.

No intensity maximum was visible at the location of IC860-a in the F606W image, and the surrounding regions are brighter than the position of IC860-a. We estimated a conservative lower limit m(606)$_{AB,IC860-a}$ $\gtrsim$22.4 by assuming the upper limit flux is equal to the square root of the mean count rate in a 0.15\arcsec radius aperture. 

Following this approach we obtained m(814)$_{AB, IC860-a}$ $\simeq$ 21.5 and the source FWHM=5.8 pixels (65~pc). The F814W IC860-a source profile is larger than the diffraction limit, noisy, possibly slightly resolved, and likely significantly affected by dust absorption. Assuming an intrinsically red color of (m(606)-m(814))$_{0,AB}$ = 0.6 for IC860-a leads to the lowest amount of dust reddening of E(m(606)-m(814))$_{AB} >$ 1.6 for A$_{\lambda} \propto \lambda^{-0.7}$. This is an extreme lower limit stemming from uncertainties in the point source photometry as shown by our dust ratio measurement in \S\ref{sec:ssctrdust} and NIR photometry by \cite{AlonsoHerrero06}.

NIR archival images of IC-860 are available from NIC2 and WFC3/IR observations. The smaller pixel scale of the NIC2 images better sample the HST PSF and are the choice for NIR photometry of the compact IC860-a source. The WFC3/IR data are more sensitive and have a wider field of view that is essential for position measurements as discussed above. We found that photometry from NIC2 and WFC3/NIR agrees to within 0.1~mag in the F160W filter (see \S\ref{sec:ic860cmz} below). Flux measurements of IC860-a are not as well determined from the WFC3/NIR data as from NIC2 due to the difference in pixel sampling, but the agreement between the NIC2 and WFC3 F160W photometry shows that IC860-a has varied by $\lesssim$0.2 ~mag during the 12.4~yr interval between these two observations. IC860-a is not a rapidly varying transient source such as a supernova.

Infrared photometry of IC860-a faces difficulties with backgrounds similar to those in the optical bands.  In our approach we estimated the photometric uncertainty based on the counts within the half maximum intensity radius. We did not include possible additional offsets due to aperture corrections as we made the conservative assumption that IC860-a contains an uncertain amount of emission from the CMZ arc. Thus our fluxes and magnitudes refer to IC860-a plus a  contribution from the CMZ, i.e., the total emission within the half width diameter of IC860-a.  

We carried out 2 pixel radius (r=0.25\arcsec) aperture photometry of IC860-a in the WFC3/IR F140W and F160W images.  The IRAF radprof task provided centered photometry with radial plots to determine the adopted backgrounds. IC860-a was cleanly detected in the archival WFC3/IR F140W and F160W images with a FWHM of 3.8 pixels in the F140W filter, slightly larger than a pure point source. Although the background was noisy, the photometry was repeatable at the 20\% level based on flux changes from small centering offsets. 

We calculated luminosities for IC860-a for D=59~Mpc, assuming $\tau_V$=3.5 with $\tau_{\lambda} \propto \lambda^{-0.7}$. To estimate the total luminosity we converted the absolute magnitudes to solar luminosities by comparing with the in-band AB absolute magnitudes of the Sun. Our preferred results came from the NIC2 F160W observation giving L(IC860-a)$ \geq$ 8 $\times$ 10$^8$~L$_{\odot}$. This is a lower limit in that our aperture did not include the full intensity profile and the total extinction is uncertain. F160W WFC3/IR photometry with an r=0.26\arcsec (2 pixel) aperture gives L(IC860-a)$_{bol}$ $\lesssim$ 2$\times$10$^9$~L$_{\odot}$. This is a strong upper limit due to the poorer image sampling that leads to inclusion of more of the CMZ.  Luminosities also depend on the quality of the foreground dust correction model, which is an additional source of uncertainty but generally leads to underestimates of fluxes and thus luminosities  \citep[see] [and the discussion below]{Gallagher24}.   We therefore adopted L(IC860-a)$_{bol} \approx$ 10$^{9\pm0.3}$~L$_{\odot}$, in agreement with the results found by \citep{AlonsoHerrero06} who derived M(160)$_{Vega}$= -19.6 corresponding to L(160)$_{IC860-a}$= 1.5 $\times$10$^9$~L$_{\odot}$. The results of the photometry are summarized in Table~\ref{tab:phot}.

A second approach to find L(IC860-a)$_{bol}$ comes from measurements of the in-band flux for each filter. Here we used the photometry key words from the image fits headers. Our most reliable results are from the better sampled NIC2 images that give in-band luminosities per filter of $\sim$10\% of 
 L$_{IC860-a,bol}$. The HST imaging data thus require that 
 L$_{IC860-a,bol}$ $>>$10$^8$~L$_{\odot}$. 

The increasing luminosity of IC860-a  with increasing wavelength in Table \ref{tab:phot} is symptomatic of the limitations of the foreground screen model that become less severe when optical depths decrease at longer wavelengths. The intensity and color data in  Table \ref{tab:phot} further demonstrate that we are not dealing with an ideal foreground dust screen, as IC860-a is both heavily reddened and has a substantial F814W surface brightness. Such a situation is not expected for a foreground screen dust model because high opacities leading to strong reddening also require low source intensities. 

If dust and luminosity sources are mixed, then highly opaque systems will emit an intensity set by the source function. The source function for pure absorption in a homogeneous medium is $S_{\lambda} = j_{\lambda}/\kappa_{\lambda}$. Here $j_{\lambda}$ is the volume emissivity that may rise into the NIR and $\kappa_{\lambda}$ is the opacity that declines with increasing $\lambda$. This case differs from the foreground screen prediction. $S_{\lambda}$ from a very optically thick system can have a significantly reddened spectral energy distribution relative to the $j_{\lambda}$ sources in combination with a substantial emergent intensity. The foreground screen model therefore provides a lower bound to the optical depths and dust obscuration corrections and our luminosities also are lower limits. For IC860-a, derived luminosities are converging in the NIR, implying that the foreground screen model may be more realistic for interpreting the F160W data, but to be conservative we tabulate all of our calculated IC860-a luminosities as lower limits. 

\begin{table}
    \caption{IC860-a photometry.}
    \centering
    \begin{tabular}{lclcl}
    \hline\hline
    \\
     Filter & m$_{AB}$ & A$_{\lambda}^a$ & log(L/L$_{\odot}$)  \\  \hline
     F606W$^b$ & $>$23.6$\pm$0.3 & $\gtrsim$3.5 & $>$7.5 \\
     F814W$^c$ & $\gtrsim$ 21.5 & $\gtrsim$2.7 & $\gtrsim$7.8 \\
     F110W$^d$  &  19.1    &   $\geq$2.1    &   $\gtrsim$8.6      \\
     F140W$^e$ & 17.9 & $\geq$1.6 & $\geq$8.9   \\
     F160W$^d$& 17.0 & $\geq$1.7 & $\geq$9.0 \\
     \hline
    \end{tabular}
     \label{tab:phot}
     \tablefoot{Photometry with r=0.15\arcsec apertures with estimated uncertainty of $\pm$0.2~mag. Aperture corrections are not included. $^a$ Dust obscuration is discussed in \S\ref{sec:ssctrdust} below. $^b$ Upper limit as IC860-a is not detected. $^c$ Data are shown as upper bounds on the observed magnitude due to modest contrast with the surroundings. This leads to lower bounds on the luminosity due to the use of the foreground screen model in combination with high optical depths. $^d$ Based on NIC2 images; F160W was used to estimate the luminosity of IC860-a; see text for discussion. $^e$ F140W data from WFC3/IR. }
\end{table}

\subsection{The IC~860 CMZ}\label{sec:ic860cmz}

We defined the IC~860 CMZ as the central high surface brightness region of $\sim$ 1.2\arcsec $\times$ 0.4\arcsec (340 $\times$ 115~pc$^2$) centered on the CO 1-0 distribution \citep{Alatalo24}. CMZ fluxes from the pair of WFC3 F140W observations separated by 6.5 months and with different HST orientations agreed to within 1\%.  We measured the observed flux ratio between the CMZ and IC860-a and found that IC860-a contributes $\approx$10\% of the total F160W flux from the CMZ. Since IC860-a appears to be more heavily obscured than the CMZ, the observed flux fraction is a lower limit to the true luminosity ratio. The results from CMZ photometry are summarized in Table~\ref{tab:cmz_phot} that gives an observed F160W mean CMZ intensity of $\mu_H$ = 14.4~magnitudes~\arcsec$^{2}$ and a  luminosity density of $\approx$  5$\times$10$^{10}$~L$_{\odot}$~kpc$^2$. The observed $\mu_H$ is near the upper bound for early-type galaxies \citep[e.g.,][]{Wu05}.  Radio continuum measurements show little flux from the CMZ aside from the area of the CON \citep{Condon90,Baan17,Aalto19}. 

\begin{figure}
   \centering
   \includegraphics[width=0.25\textwidth]{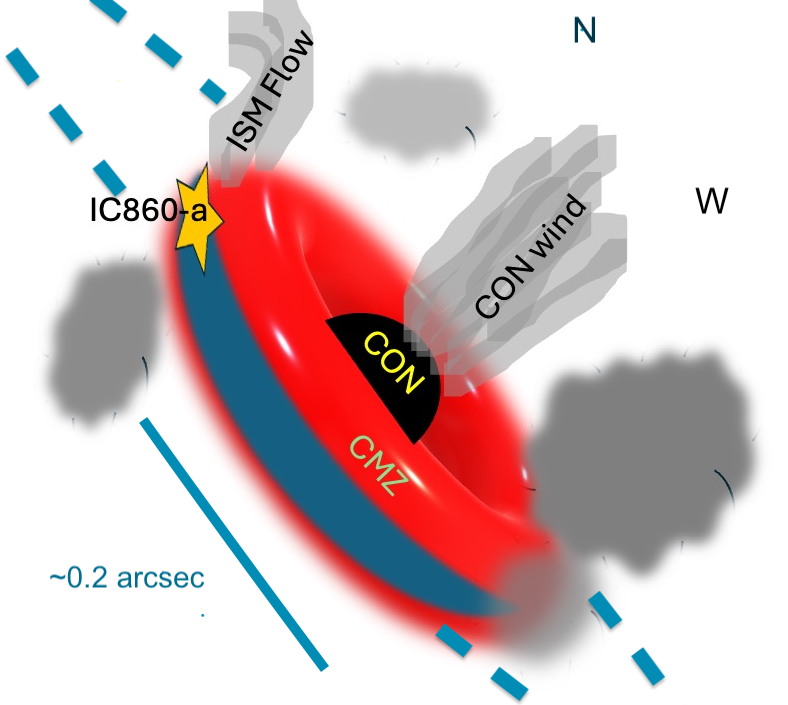}
      \caption{Conceptual cartoon view of the inner 100~pc of the IC~860 CMZ as observed in the NIR. The CON, spherical in this idealization, is opaque in the NIR, while the inner high brightness annular zone of the CMZ is illuminated with a peak intensity near its edge. In this part of the CMZ a mixture of stars and dust leads to a combination of substantial reddening and high intensity in the NIR. This situation does not prevail at the optically thick surface of the CON that we observe.  The sharp edge of the inner CMZ in this cartoon is not physical. The dashed lines show that the CMZ extends well beyond the illustrated central region. The NIR intensity and foreground optical depth estimates peak at the position of IC860-a, an indication that dust is associated with IC860-a. The area to the right (west) of the CON has higher levels of dust obscuration, likely due to dusty outflows. Dust structures with moderately high opacity around IC860-a and elsewhere are suggestive of additional gas flows.}
         \label{fig:ic860_cmz_model}
\end{figure}

The molecular ISM-rich CMZ that is luminous in the NIR surrounds the CON. In IC~860 the CON is a central peak above a relatively smooth, larger-scale radial gradient in the density of interstellar matter. The absence of NIR emission from the CON requires that the distribution of its luminosity sources relative to dusty gas differs in a major way from that in the CMZ. The source function in the CMZ leads to substantial NIR intensities while NIR emission from the CON is not detected. We can understand this change if most luminosity sources in the CON are located behind a $\tau_{NIR} >$ several dust screen.   Figure \ref{fig:ic860_cmz_model} is a cartoon model of the R$\lesssim$100~pc center of the IC~860 CMZ based on our qualitative analysis.

 We made a Pa$\alpha$ emission line map (Figure~\ref{fig:nic2_cmz}, lower right corner) by scaling the NIC2 F187N filter image as an off band to subtract the continuum from the F190N image that contains the Pa$\alpha$ emission line (see Figure~\ref{fig:nic2_cmz}).  The peak of the Pa$\alpha$ emission is slightly offset from the F190N continuum peak of IC860-a with the Pa$\alpha$ emission being more spatially extended than the NIR continuum. \cite{AlonsoHerrero06} interpreted the presence of Pa$\alpha$ emission along much of the CMZ arc as evidence for the presence of young massive stars within the CMZ and IC860-a. However, we have yet to determine whether the Pa$\alpha$ emission is due to stellar photoionization. We defer a discussion of properties of emission lines and possible levels of associated star formation activity in the IC~860 CMZ to the analysis of JWST spectra obtained in program JWST~GO-01991, P. I. G. Privon that can better constrain its properties (Privon  \emph{et al.} in prep.).

\begin{table}
    \caption{IC860-a CMZ infrared photometry.}
    \centering
    \begin{tabular}{llcll}
    \hline\hline
    \\
     Filter & m$_{AB}$ & A$_{\lambda}^a$ & log(L/L$_{\odot}$$^b$) \\  \hline
     WF3-F140W & 15.4 & $\geq$1.4 &  $\sim$9.8  \\
     WF3-F160W & 15.2  & $\geq$1.3 &  $\sim$9.8  \\
     NI-F160W   & 15.3     &  $\geq$1.3    &  $\sim$9.8  \\
       \hline
    \end{tabular}
     \label{tab:cmz_phot}
     \tablefoot{ CMZ defined by photometry with the isophote shown in Figure~\ref{fig:ic860_opt} corresponding to $\mu_{AB}$=16.6~mag\arcsec$^2$ in the WFC3 F140W filter. $^a$Foreground screen extinction scaled from an assumed mean $\tau_V$=2.5 for the CMZ (see \S\ref{sec:ssctrdust} below). $^b$Luminosity based on the tabulated A$_{\lambda}$. As discussed in \S\ref{sec:ssctrdust}, similar luminosities in the NIR suggests that the foreground screen approximation is providing reasonable results.}
\end{table}

\section{Central dust obscuration}\label{sec:ssctrdust}

\begin{figure}
   \centering
\includegraphics[width=0.45\textwidth]{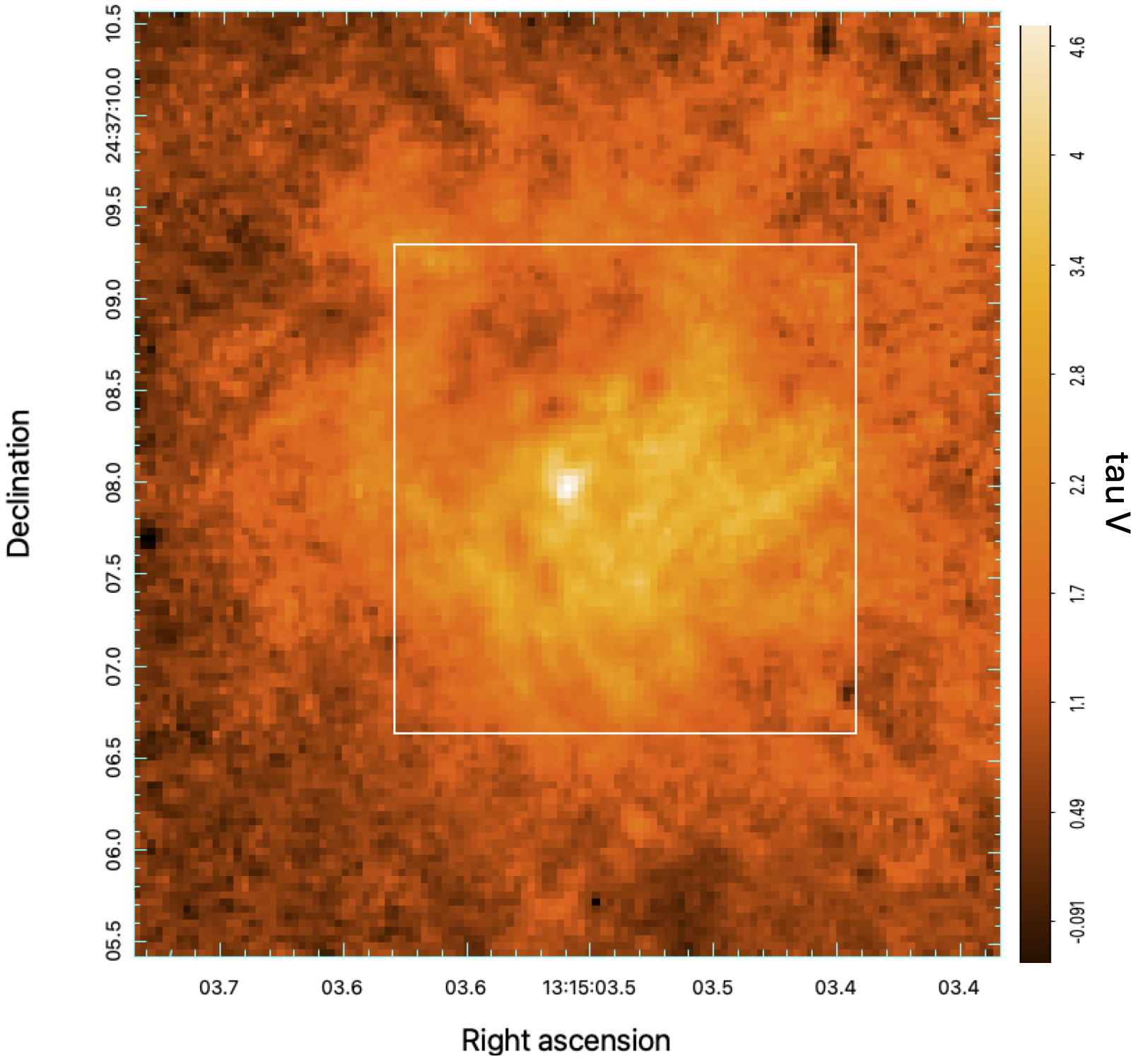}
\includegraphics[width=0.20\textwidth]{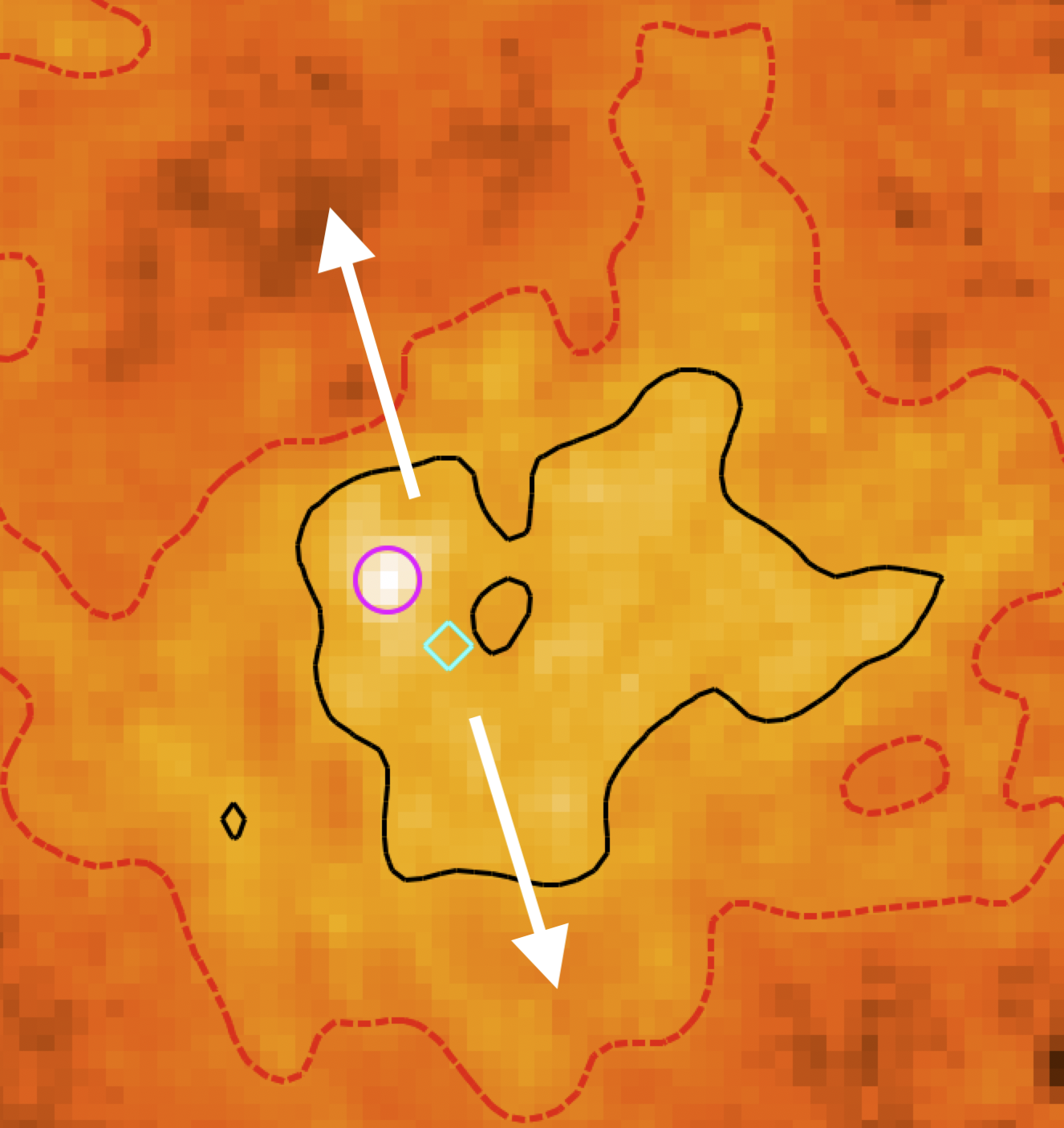} 
      \caption{{\it Upper} Estimated V-band optical depths in the central region of IC~860. The value of $\tau_V$ was calculated from the m(606)$ - $m(814) HST color relative to the blue circumnuclear color of m(606)$ - $m(f814) = -0.3.  {\it Lower} Zoomed-in image  of the CMZ in the 2.5 $\times$ 2.7~arcseccond$^2$ (0.7 $\times$ 0.8~kpc$^2$) area shown by the white box.  The dashed red contour is for $\tau_V$=2.0 and the black contour marks $\tau_V$=3.0. A circle marks IC860-a, the diamond is at the position of the CON, and the arrows point along the major axis of the CMZ.}
         \label{fig:dustdepth}
   \end{figure}

Dust features in the center of IC~860 dominate its optical appearance as shown in Figure~\ref{fig:ic860_opt} (see \cite{Luo22} for discussions of the larger scale dust distribution). We produced the dust visual optical depth map of IC~860 in Figure~\ref{fig:dustdepth} from the ratio of count rates in the F606W to those in the F814W WFC3 images. Optical depths assume a purely absorbing foreground dust screen. We followed the methodology based on intensity ratios in dusty regions compared to low obscuration regions as described by \cite{Gallagher24} and \cite{Schisgal25}. As in these previous studies, we used a power law form for the long wavelength obscuration, $\tau_{\lambda}= \tau_V(\lambda/550~nm)^{-0.7}$
This technique.\footnote{\cite{AlonsoHerrero06} adopted a power law with an exponent of -0.8}  provides lower limits to the calculated optical depths presented in Figure~\ref{fig:dustdepth}. The CON is observed in the NIR as a region that is opaque at the wavelengths of the HST observations. It does not contribute to color maps and the opacity estimate from the foreground dust screen model is not affected by the CON.  As discussed by \cite{Gallagher24}, this combination of features demonstrates that high opacity originates within the CON  rather than arising in more extensive, less dense foreground screen of interstellar matter. 

\begin{figure}
	\centering
\includegraphics[width=0.45\textwidth]{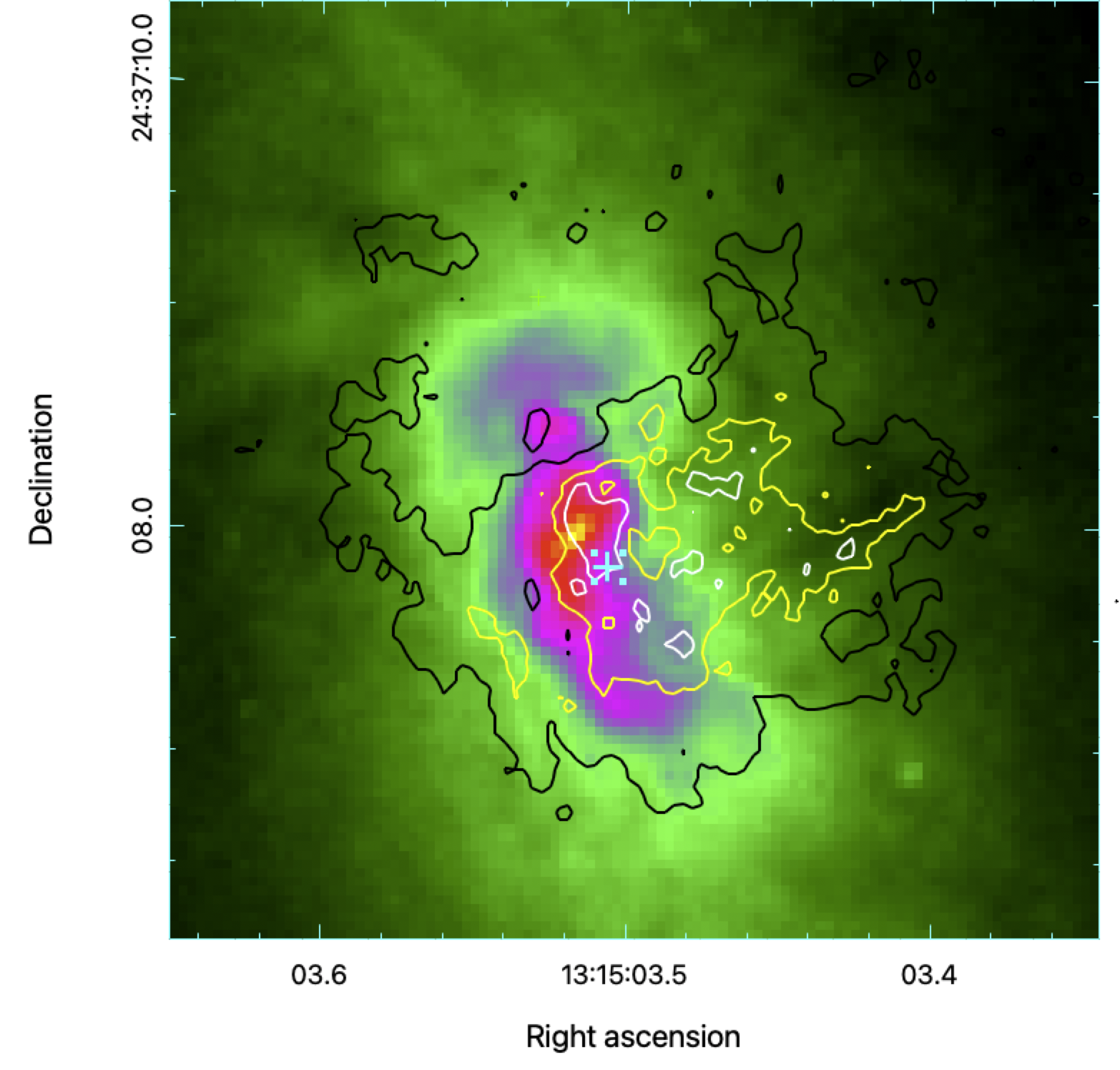}
	\caption{False color version of the F814W image of the center of IC 860. This figure shows the brightness distribution with foreground $	\tau_V$ contours superimposed. IC860-a is the compact spot above and slightly to the left (to the north-northeast) of the CON. The contour levels for $\tau_V$ are white=3.5 yellow=3.0, and black=2.0. Very substantial reddening is present across the $\sim$3 $\times$10$^5$~pc$^{2}$ central region. The cross marks the location of the CON. The distribution of dust extending above the disk of IC~860 shows that outflows occur across the CMZ, similar to the situation in the CON-host Zw~049.057 \citep{Gallagher24}.}
\label{fig:dustmap}
\end{figure}

As \cite{Luo22} discuss, most of the center of IC~860 is highly obscured with $\tau_V \geq$2. The dust absorption mainly extends to the west, approximately perpendicular to the major axis of IC~860. The high opacity central dust screen in IC~860 is best understood as originating from out-of-plane gas flows, consistent with observations of central molecular flows \citep[e.g.,][]{Aalto15,Luo22,Gorski23}. Central regions of enhanced brightness visible in the F438W data in Figure~\ref{fig:ic860_opt} have bluer colors than their surroundings. They are likely to be areas containing young stars observed through patches of low dust obscuration. However, we did not model these regions since they do not align with either the CON or IC~860-a that are the focus of this study.

An outstanding feature of our HST F606W--F814W flux-ratio color map is the compact high opacity feature that is coincident with IC860-a and offset from the location of the CON. As we discuss in \S\ref{sec:northobjptm}, detection of IC860-a in the F814W image and its absence the F606W image naturally leads to the associated compact, deep dust absorption feature that is only slightly more extended than the F814W point spread function.  We photometered IC860-a in the color-ratio image with an r=0.079\arcsec (2 pixel) circular aperture corresponding to a diameter of approximately 40~pc. We adopted I$_{F606W}$/I$_{814W}$=1.65$\pm$0.05 for the unobscured galaxy background color-ratio taken from the surrounding regions with low dust obscuration. Our aperture photometry yielded a formal foreground screen dust opacity estimate for IC860-a of $\tau_V \gtrsim 3.5\pm$0.5 within the 2 pixel radius aperture.   As shown in Figures~\ref{fig:dustmap} and \ref{fig:dustdepth} this reddening maximum is near the boundary of the white and yellow contours that are part the dust complex extending over much of the central region of IC~860. Although measurements of $\tau_V$>3 are necessarily uncertain (as well as being lower limits), our HST V-I color ratio map indicates that IC860-a is substantially more deeply obscured than its surroundings and so likely contains dusty interstellar matter. 

The CON is located near the lower (southern) edge of the white contour in Figure~\ref{fig:dustmap} and within the most obscured contour in Figure~\ref{fig:dustdepth}, a position that is not unique in terms of  the central dust obscuration. HST NIR and optical observations of the centers of the CON-LIRGs NGC~4418  and Zw~049.057 do not clearly reveal the CONs as discrete sources of NIR emission or absorption \citep{Scoville00,Gallagher24,Schisgal25}. In these galaxies, as in IC~860, the CON's, while in dust obscured galaxy centers, are not positioned where the local dust foreground shows signs of unusually high optical depths. 
 
Our observed $\sim$20~pc radius of the NIR  absorption from the IC~860 CON is $\geq$2 times the R$_{submm}$, which follows the pattern expected from the radiative transfer model by \cite{GonzalezAlfonso19} for IC~860. This transfer model adopts a spherical structure, radius of 28~pc, gas column of  N(H)$\approx$10$^{25}$~cm$^{-2}$, inward, radially increasing dust densities, and either young stars or AGN as luminosity sources. It  predicts, as we observed, that the submillimeter emission core of the CON is embedded in a somewhat larger envelope of dense interstellar matter. We adopted a spheroidal CON model for this paper (see Eq. (1) below) as the simplest option, an assumption that needs to be confirmed.

\section{Discussion}\label{sec:discuss}
\subsection{The IC~860 nuclear region }

The two compact central luminosity sources in IC~860 contribute in distinct ways to the central luminosity of the galaxy. Our analysis in combination with the \cite{AlonsoHerrero06} study prove that IC860-a and the CMZ produce the central region's NIR luminosity while submillimeter and radio observations show that the CON dominates in the total luminosity in the far-IR. This combination of wavelength-dependent contributions complicates previous interpretations of the mid-IR spectra of IC~860, and optical-NIR luminosity determinations. 

Properties derived from observations of the center of IC~860 are strongly dependent on the wavelength coverage and angular resolution of the observations. For example, \cite{Lahuis07} analyzed {\emph Spitzer Space Telescope} (SST) mid-infrared spectra of IC~860 taken with a 4.7\arcsec width slit. The SST instrumental point spread function in the 14~micron region of interest is about 3.5\arcsec. The SST spectrograph slit therefore encompassed the entire CMZ and its immediate surroundings, sampling emission from the CMZ, IC860-a, and inner bar, as well as any mid-infrared emission from the CON.  

\subsection{Masses}

The presence of  IC860-a as a compact source in the inner IC~860 CMZ raises questions about its relationship, if any, to the CON. Is the IC~860 CON an unusual example of a galaxy nucleus or could it be an extreme version of a star-forming molecular cloud as in the Medusa galaxy \citep{Koenig14}? Unlike the Medusa's molecular complex, the CON is at the centroid velocity of IC~860, is at the center of the CMZ, is surrounded by rotating gas disks with regular structures with velocity fields that are aligned in position angle with the main stellar disk, and is the radio continuum point source \citep{Baan17,Aalto19,Luo22}. These properties are consistent with expectations for the CON being the nucleus of IC~860. 

In principle, the mass of the CON can distinguish whether it is a nucleus.  In a young star-forming cloud with the normal low star formation efficiency,  the total mass will be close to that of the molecular medium. A nuclear CON  should also include substantial mass contributions from preexisting stars and likely a central supermassive black hole. We also would normally expect the mass of the nucleus of IC~860 to exceed that of any compact circumnuclear stellar complexes that form in the CMZ.

\subsection{Mass of the CON}

Unfortunately, dynamical mass estimates for the IC~860 CON are difficult to obtain due to uncertainties in orientation and internal dynamics. Mass estimates range between 2$\times$10$^7$~M$_{\odot}$ and $\lesssim$9 $\times$10$^7$~M$_{\odot}$ within R=10~pc depending on the inclination of the rotating component of molecular gas within the CON \citep{Aalto19}. We prefer the higher CON mass that assumes an inclination of 60$^{\circ}$ that fits with the flattening derived from the CMZ's isophotes. The molecular mass in the CON can be crudely estimated by assuming a spheroidal distribution of interstellar matter with radius R$_{dust}$  corresponding to the assumed gas column density N$_H$,
\begin{equation}
   \rm{M_{ISM}}/M_{\odot}\sim 4 \times 10^{7} M_{\odot}(N_H/(10^{25}cm^{-2}(R_{dust}/20~pc)^2\epsilon_{CON},
\end{equation}
where $\epsilon_{CON}$ $\leq$1.0 is a parameter to account for the possible  flattening of the CON's gaseous envelope. 

The significant optical depth at the R$_{NIR} \sim$ 20~pc absorption edge of the CON provides only a weak limit of N$_H$(20~pc)$\gtrsim$10$^{22}$~cm$^{-2}$. The value of N(H) for the CON thus rests on the model-dependent analysis of molecular line observations. For this study we adopted N$_H$=10$^{25}$~cm$^{-2}$ from the greenhouse models by  \cite{GonzalezAlfonso19}.  The NIC2 observations, as discussed in \S\ref{sec:centralstruct}, suggest a radius R$_{dust}$ $\approx$ 20~pc for the high NIR opacity outer limit of the CON that we adopted to estimate the total dynamical mass of the CON. \cite{Aalto19} also find the molecular CON is rotating with a $\sigma$/V$_{rot} \approx$0.5 implying that the CON is rotationally flattened so that $\epsilon_{CON} <$ 1. Our data lack the resolution to determine if the CON is flattened at R$_{dust}$=20~pc. Therefore we simply extrapolated the \cite{Aalto19} dynamical model to the CON mass enclosed within 20~pc with a constant V$_{rot}$ that gives \rm{M$_{dyn}$} $\sim$2 $\times$ 10$^8$~M$_{\odot}$ implyng that a significant fraction of its mass in the form of gas if $\epsilon_{CON}$ is not small. 

This combination of parameters, despite their considerable uncertainties, limits the stellar and central black hole mass within the R$_{NIR}\approx$20~pc of the CON to $\lesssim$10$^{8}$~M$_{\odot}$. Refining mass measurements for the CON would help to constrain the CON's internal structure. Nuclear star clusters and their surroundings in the 10$^{7-8}$~M$_{\odot}$ mass range primarily are found in late-type spirals \citep[e.g.,][]{Kormendy93,Barth09,Georgiev16}. The Milky Way galaxy with (log(\rm{M$_{\star}$}) $\approx$ 10$^{10.8}$~M$_{\odot}$ offers a point of comparison.  The mass within a radius of 20~pc from the Milky Way's nucleus is $\sim$5 $\times$ 10$^7$~M$_{\odot}$ \citep[e.g.,][]{Feldmeier-Krause25}. Thus if the Milky Way were to add the gas cloak of a CON, we would expect its total mass to be $\sim$10$^8$~M$_{\odot}$, similar to the situation in IC~860.  Unfortunately,  the current uncertainties in the mass of the IC~860 CON are 
too large to establish whether the CON is a heavily dust obscured version of a normal galaxy nucleus.

\subsection{Mass of IC860-a}

The mass of IC860-a can be estimated from its luminosity by adopting a stellar model and making an assumption about the nature of this source. In considering the possibility that it is a galaxy nucleus, we assumed that IC860-a is not a radio-quiet AGN based on its lack of X-ray emission \citep{Luo22}. We derived a stellar mass for a nucleus model by adopting an old stellar population NIR mass-to-light ratio of $\gamma_{160}$=1.0 where the central black hole makes a minor contribution to the mass within R$\approx$30~pc.
For a young stellar complex model we take $\gamma_{160}$=0.3 that would normally be associated with an actively star-forming system. A significantly lower value of $\gamma_{160}$ requires an unusual bottom-light or top-heavy stellar initial mass function. The mass of IC860-a for a stellar model with a Kroupa-type initial mass function is
\begin{equation}
\rm{M}_{IC860-a,*} \approx 3 \times 10^8 (L_{IC860-a,bol}/10^9~L_{\odot})(\gamma_{160}/0.3)~M_{\odot}. 
\end{equation}
 
Given the concentration of molecular gas in the center of IC~860, a young, massive star-forming complex with R $\approx$ 30~pc could explain IC860-a. However, the photometric mass of \rm{M}$_{IC860-a,*} \sim 3 \times$10$^8$~M$_{\odot}$ for such a system is extremely high for a compact star-forming region. Furthermore, we are not aware of observations showing that IC860-a contains a significant amount of molecular gas. While the mass estimate for such a young region could be reduced if additional light contributions come from the CMZ, a young stellar explanation for IC860-a requires an unusual event. A further complication arises in that the mass for a young stellar complex model for IC~860-a is potentially similar to the mass estimated for the IC~860 CON. If IC860-a is a young stellar complex, then its presence would demonstrate that star formation can be a significant gas sink within the CMZ that competes with the CON for interstellar material. 

A different possibility is that IC860-a is an inactive galaxy nucleus, possibly a remainder from a past merger \citep[see discussion of a merger in][]{Luo22}. This option leads to some concerns but cannot be excluded with the present data. For a nucleus, eq.(2) with $\gamma_{160}$ = 1.0 leads to an inferred stellar mass of $\gtrsim$10$^9$~M$_{\odot}$ for IC860-a which would exceed the estimated mass within the same radius of the CON and is $\sim$10-20\% of the  $\sim$10$^{10}$~M$_{\odot}$ mass of the CMZ.  The lack of major dynamical anomalies in the CMZ CO 1-0 velocity field associated with  IC860-a \citep{Luo22,Alatalo24} implies that if IC860-a orbits within the CMZ, it is significantly less massive than the CMZ.  Dynamical modeling could determine limits on the mass ratio between an IC860-a nucleus and CMZ ithat would avoid detectable perturbations of the  kinematics of  the CMZ CO disk.  Alternatively, perturbations of molecular gas in the CMZ  by IC860-a could be minimized if it is located out of the plane of the IC~860 CMZ. The heavy dust extinction might result from IC860-a being behind the galaxy, seen through the dusty edge of the CMZ. However, this may be inconsistent with the JWST detection of mid-infrared thermal dust emission from IC~860-a (Privon {\it et al.} in prep.) that suggests dust is physically associated with IC860-a. More sensitive measurements of the structure and kinematics of the CON, IC860-a, and the CMZ are needed to constrain the properties of this complex region. 

\section{Conclusions}\label{sec:conclude}

The LIRG IC~860 contains a pair of compact, luminous central sources: The CON is observed as an opaque central region in the NIR but provides the bulk of the system's luminosity that emerges in the far infrared.  IC~860-a is a compact, prominent NIR source with properties consistent with either a massive young stellar complex or second galaxy nucleus.  Both of these objects are at projected positions within the CMZ. 

\begin{itemize}

\item This is the first identification of a CON as an opaque feature in the NIR. The NIR extent of absorption due to the CON is approximately 20~pc, $\gtrsim$ 2 times the radius of the CON when observed as a submillimeter source. In \S\ref{sec:ssctrdust}, we established that this NIR CON radius when interpreted with a spheroidal model is in reasonable agreement with greenhouse CON having an N(H) column similar to the values observed for the IC~860 CON. The NIR properties of the CON therefore are consistent with the presence of a massive concentration of low angular momentum molecular gas. The lack of highly reddened NIR emission from the CON indicates that its power sources lie behind a dust layer that is substantially optically thick at the  2~$m\mu$ long wavelength limit of the HST data.

\item IC860-a originally was identified by \cite{AlonsoHerrero06} as the IC~860 nucleus. We confirm their finding that it is heavily dust obscured with $\tau_V \gtrsim$ 3.5. IC860-a is offset by $\approx$50~pc (0.16$\pm$0.03\arcsec) in projection from the IC~860 CON that has appears to be the nucleus of IC~860. IC~860 contains two luminous central compact sources. IC~860-a, with L$\sim$10$^9$~L$_{\odot}$ which has not varied significantly over 12.4~yr and makes a modest contribution to the central NIR luminosity. and the nuclear CON that provides the bulk of the $\sim$10$^{11}$~L$_{\odot}$ total power radiated in the FIR from IC~860. 

\item Within R$\lesssim$ 100~pc IC~860 is blanketed by a complex distribution of   extraplanar dust with optical depths of $\tau_V \gtrsim$ 2. Our dust opacities are based on foreground screen models and likely underestimate the true obscuration levels; our luminosities based on photometric measurements also may be underestimates.  The existence of the extended extraplanar interstellar medium traced by dust absorption is consistent with molecular line evidence for large spatial scale gas flows, and especially with the detection of outflows and inflows associated with the CON. 
    
\item A disk like structure exists in the center of the IC~860 bar. We identify the innermost part of  this zone as an inclined disk-like CMZ with R$\sim$ 70~pc. In the NIR brighter edge of the CMZ is inclined toward us with an R$_{opaque} \approx$ 20~pc deeply obscured core at the location of the CON. Pa$\alpha$ emission is present along the rim of the inner CMZ arc, indicating that ionization sources, uncertain in nature but possibly young stars, are associated with the optically thick CMZ. 
    
 \item Our photometric stellar mass estimate for IC860-a is M$\gtrsim$ 10$^{8-9}$~M$_{\odot}$. The lower mass range for IC860-a is similar to dynamical mass estimates for the IC~860 CON of $\sim$10$^8$~M$_{\odot}$ within R=20~pc, which is typical of a moderate mass galaxy nucleus. 
    
\item The physical nature of IC~860-a is not clear. If IC860-a is an intruding second nucleus, then its mass likely exceeds that of the CON. This introduces issues stemming from the lack of evidence for perturbations of the regular kinematics of the potentially nearby molecular gas disk in the CMZ. If IC860-a is a young stellar complex, then its inferred mass of $\sim$10$^8$~M$_{\odot}$, is unusually high  and possibly comparable to the mass of the CON. The nature of this unusual object and its impact on the evolution of the CMZ and CON remain to be determined.
    
 \item The presence of two massive compact objects in the center of IC~860 within a 50~pc projected distance complicates efforts to understand the origins of the CON. A second nucleus would be suggestive of a ULIRG-like CON formation process involving nucleus-nucleus interactions. The presence of a massive young stellar complex could indicate that the CON has had significant competition for obtaining gas from the CMZ. If IC860-a is a young stellar complex, then both it and the CON are products of the extremely gas-rich environment which is accessible to more detailed studies. Better determinations of the properties of IC860-a, the CON, and the kinematics of the inner CMZ are essential for modeling the current conditions and evolution of IC~860 and its CON.

\end{itemize}

\begin{acknowledgements}
Based on observations made with the NASA/ESA Hubble Space Telescope, and obtained from the Hubble Legacy Archive, which is a collaboration between the Space Telescope Science Institute (STScI/NASA), the Space Telescope European Coordinating Facility (ST-ECF/ESAC/ESA) and the Canadian Astronomy Data Centre (CADC/NRC/CSA). This study is associated with program JWST~GO-01991. Support for this research as a part of program JWST~GO-01991 was provided by NASA through a grant from the Space Telescope Science Institute, which is operated by the Association of Universities for Research in Astronomy, Inc., under NASA contract NAS~5-26555. The National Radio
Astronomy Observatory and Green Bank Observatory are facilities of the U.S. National Science Foundation operated under cooperative agreement by Associated Universities, Inc. JG, LS, and WStJ thank Macalester College for providing computing resources for this research. We also express appreciation to STScI and the NICMOS and WFC3 instrument teams for providing the excellent archival observations of IC~860, the CON-quest team for the many fruitful discussions of CON properties, and Dr. Elena Sabbi for her advice on the procedures for assigning celestial coordinates to HST images. This paper was improved through consideration of issues identified in a constructive report by an anonymous referee.

{\it Software:} SAOimage DS9, IRAF, Python
\end{acknowledgements}

\bibliographystyle{aa.bst} 
\bibliography{aa58938-26cor_jg2.bib} 
\end{document}